\documentclass[final, times]{elsarticle}
\usepackage[letterpaper,top=2cm,bottom=2cm,left=3cm,right=3cm,marginparwidth=1.75cm]{geometry}
\usepackage{amssymb}
\usepackage{lipsum}
\usepackage{amssymb}
\usepackage{amsmath}
\usepackage{color}
\usepackage{xspace}
\usepackage{caption}
\usepackage{appendix}
\usepackage{float}
\usepackage{hyperref}
\usepackage{diagbox}
\usepackage{soul}
\usepackage{slashed}
\usepackage{amsfonts}
\usepackage{multirow}
\usepackage{mathrsfs}
\usepackage{graphicx}
\usepackage{bm}
\usepackage{bbm}
\usepackage{comment}
\usepackage{cuted}
\usepackage{setspace}
\biboptions{sort&compress}

\newcommand{\nn}{\nonumber}

\begin{document}

%\tnotetext[]{Preprint: arXiv:2508.12345}
%\journal{Preprint CPTNP-2025-028}

\begin{frontmatter}

\title{ QCD corrections to Higgs boson pair production and decay \\
to the  $b\bar{b}\tau^+\tau^-$ final state }

\author[a]{ Hai Tao Li}
\author[a]{ Zong-Guo Si}
\author[a,b]{ Jian Wang}
\author[a]{ Xiao Zhang}
\author[a]{ Dan Zhao}
\address[a]{School of Physics, Shandong University, Jinan, Shandong 250100, China}
\address[b]{Center for High Energy Physics, Peking University, Beijing 100871, China}

\begin{abstract}
We present a comprehensive investigation of next-to-leading order (NLO) quantum chromodynamics (QCD) corrections to Higgs boson pair production and the subsequent decay into $b\bar{b}\tau^+ \tau^-$.  
We adopt the narrow-width approximation to separate the production and decay contributions and employ the dipole subtraction method to address infrared divergences inherent in perturbative QCD corrections. 
After applying a set of typical experimental cuts,
we show that NLO QCD corrections to the decay process induce a significant reduction of the fiducial cross section and reshape important kinematic distributions at the 13.6 TeV LHC, such as the invariant mass of the Higgs boson pair and the transverse momentum of the leading $b$-jet. 
We also investigate the dependence of the kinematic distributions on the Higgs self-coupling and provide the signal acceptance as a function of the Higgs self-coupling modifier with full QCD corrections.
\end{abstract}

\end{frontmatter}

\section{Introduction}

As the last elementary particle discovered at the LHC in 2012~\cite{ATLAS:2012yve, CMS:2012qbp}, the Higgs boson plays a crucial role in the Standard Model (SM) and various scenarios beyond the SM. Current measurements of its properties, including mass, width, spin, and CP property, are consistent with the SM predictions~\cite{CMS:2020xrn,ATLAS:2023zhs,CMS:2022ley,ATLAS:2015zhl,CMS:2014nkk}. The spontaneous electroweak symmetry breaking mechanism implies that the couplings of the Higgs boson with other particles are proportional to their masses, a relationship that has been experimentally validated by the ATLAS and CMS Collaborations~\cite{ATLAS:2022vkf,CMS:2022dwd}. Despite these achievements, the self-couplings of the Higgs boson remain largely unconstrained, leaving an important gap in our understanding of the Higgs sector \cite{Adhikary:2022zcu,deBlas:2019rxi}.

Higgs boson pair production serves as a direct probe of the Higgs trilinear self-coupling $\lambda_{\rm HHH}$. 
At the LHC, Higgs boson pairs are predominantly produced via gluon-gluon fusion (ggF) and vector boson fusion (VBF).  Experimentally, the Higgs boson pair events have been searched for through various decay channels, including $b\bar{b}\gamma\gamma$, $b\bar{b}\tau^+\tau^-$, $b\bar{b}b\bar{b}$, $b\bar{b}ZZ$, and multilepton final states by CMS ~\cite{CMS:2020tkr,CMS:2022hgz,CMS:2022cpr,CMS:2022gjd,CMS:2022kdx,CMS:2022omp}, as well as by ATLAS in the $b\bar{b}\gamma\gamma$, $b\bar{b}\tau^+\tau^-$, $b\bar{b}b\bar{b}$, $b\bar{b}WW^*$, $WW^*\gamma\gamma$ and $WW^* WW^*$ decay channels \cite{ATLAS:2021ifb,ATLAS:2022xzm,ATLAS:2023qzf,ATLAS:2018fpd,ATLAS:2018hqk,ATLAS:2018ili,ATLAS:2024pov}. 
Current experimental constraints on the Higgs self-coupling modifier $\kappa_\lambda \equiv \lambda_{\rm HHH}/\lambda_{\rm HHH}^{\rm SM}$  from the single- and
double-Higgs production lie in the range of $(-0.4, 6.3)$  and $(-1.2, 7.5)$ by ATLAS \cite{ATLAS:2022jtk} and  CMS \cite{CMS:2023cee}, respectively, at 95\% confidence level, 
assuming that the other couplings are the same as the SM values. Future runs of the LHC, particularly the High-Luminosity LHC (HL-LHC), are expected to greatly refine these measurements \cite{ATLAS:2025wdq}.

On the theoretical front, significant progress has been made in providing precision predictions on the production cross sections. For the ggF process, calculations have been performed up to QCD next-to-next-to-next-to-leading order ($\rm N^3LO$) \cite{Dawson:1998py,deFlorian:2013jea,deFlorian:2016uhr,Chen:2019lzz,Chen:2019fhs} and next-to-next-to-next-to-leading logarithmic ($\rm N^3LL$) level \cite{Shao:2013bz,deFlorian:2015moa,Ajjath:2022kpv} in the heavy top-quark limit, as well as to QCD next-to-leading order ($\rm NLO$) \cite{Borowka:2016ehy, Borowka:2016ypz, Baglio:2018lrj, Baglio:2020ini} and electroweak (EW) NLO \cite{Borowka:2018pxx,Muhlleitner:2022ijf,Davies:2022ram,Davies:2023npk,Bi:2023bnq,Heinrich:2024dnz,Zhang:2024rix,Li:2024iio,Davies:2025wke,Bonetti:2025vfd} with full top-quark mass dependence. Similarly, the VBF production mode has been computed up to QCD $\rm N^3LO$ \cite{Frederix:2014hta,Ling:2014sne,Dreyer:2018qbw,Dreyer:2018rfu,Jager:2025isz} and EW NLO~\cite{Li:2024iio,Dreyer:2020xaj}.
Given that the Higgs boson pair production rate is rather small, it is almost mandatory that experimentalists search for the signals with at least one of the Higgs bosons decaying into bottom quarks, which has the largest branching fraction.
In such a case, the QCD corrections in $H\to b\bar{b}$ decay cannot be neglected. 
Indeed, we have demonstrated that QCD NLO corrections to the decay process in $gg\to HH \to b\bar{b} \gamma\gamma$ result in substantial effects on both the inclusive and differential cross sections, which significantly exceed the N$^3$LO QCD correction to the production process in the heavy top-quark limit \cite{Li:2024ujf}.

In this work, we extend the previous study to $gg\to HH \to b\bar{b} \tau^+\tau^-$, presenting a full calculation of NLO QCD corrections in both the production and decay processes.
This channel is especially appealing because of its relatively large branching fraction, around $7.3\%$, which enhances the overall signal yield compared to other decay modes, such as $b\bar{b}\gamma\gamma$. 
Moreover, despite the experimental challenges associated with $\tau$ lepton identification, the $b\bar{b}\tau^+\tau^-$ final state offers distinctive kinematic signatures that improve the discrimination ability between signal and backgrounds. 
Similar to the $b\bar{b}\gamma\gamma$ channel~\cite{Li:2024ujf}, we find that, after applying typical kinematical cuts, the QCD corrections to the decay lead to a significant reduction of the fiducial cross section as well as remarkable reshaping of key kinematic distributions.
The signal acceptance is essential to extracting constraints on the total cross section from the observed events under kinematic cuts.
We provide the signal acceptance in this process as a function of $\kappa_{\lambda}$.

The rest of this paper is organized as follows. In section~\ref{sec:frame} we present the framework for QCD corrections to the production and decay of the Higgs boson pair. The numerical results are discussed in section~\ref{sec:num}. We conclude in section~\ref{sec:concl}.

\section{Theoretical framework}
\label{sec:frame}

We are going to calculate NLO QCD corrections to Higgs pair production and decay, $gg\to HH\to b\bar{b}\tau^+\tau^- $.
Because the Higgs boson width ($\Gamma_H=4.1$ MeV) is so small compared to its mass ($m_H=125$ GeV)~\cite{ParticleDataGroup:2024cfk}, 
the on-shell production gives the dominant contribution to the cross section.
Moreover, there is no interference between the corrections to the production and decay at NLO in QCD, since the Higgs boson is colorless.
Therefore, the cross section can be written in the narrow width approximation as
\begin{align}\label{eq:crxs}
  \int d \sigma_{\rm pro+dec} & =  \int d \sigma_{\rm pro} (gg\to H_1 H_2)\times \frac{\int d\Gamma_{H_1\to b\bar{b}}}{\Gamma_{H}} \frac{\int d\Gamma_{H_2\to \tau^+ \tau^-}}{\Gamma_{H}} \nonumber\\
&   =  \int d \sigma_{\rm pro} (gg\to HH) \times \frac{\int d\Gamma_{H_1\to b\bar{b}}}{\Gamma_{H\to b\bar{b}}} \frac{\int d\Gamma_{H_2\to \tau^+ \tau^-}}{\Gamma_{H\to \tau^+ \tau^-}} \times  B_r\left(b\bar{b}\tau^+\tau^-\right)~,
\end{align}
where $\Gamma_{H}$ is the total decay width and $\Gamma_{H\to X}$ is the partial decay width of $H\to X$. 
In the first line, we have denoted the Higgs bosons decaying to $b\bar{b}$ and $\tau^+\tau^-$ as $H_1$ and $H_2$, respectively, which are not considered as identical particles. 
In the second line, we adopt the conventional notation for the production of identical Higgs bosons.
Consequently, $B_r\left(b\bar{b}\tau^+\tau^-\right)$ is equal to twice the product of the branching ratios of $H\to b\bar{b}$ and $H\to \tau^+\tau^-$. 
It appears in the formula as an overall factor and has been computed very precisely \cite{LHCHiggsCrossSectionWorkingGroup:2016ypw}.
Moreover, it can be measured directly.
Therefore we take it as a fixed quantity in our calculation\footnote{The same strategy has been adopted in the calculation of $pp\to W(l\nu)H(b\bar{b})$ \cite{Caola:2017xuq}.} , 
while the other parts of Eq.~(\ref{eq:crxs}) are expanded in the strong coupling $\alpha_s$.

From Eq.~(\ref{eq:crxs}), the LO cross section of $gg\to HH\to b\bar{b}\tau^+\tau^- $ is given by 
\begin{align}  
  \int d \sigma_{\rm pro+dec}^{\rm LO} =   \int d \sigma_{\rm pro}^{\rm LO} 
    \times\frac{\int d\Gamma_{H_1\to b\bar{b}}^{\rm LO}}{\Gamma_{H\to b\bar{b}}^{\rm LO}} \frac{\int d\Gamma_{H_2\to \tau^+ \tau^-}}{\Gamma_{H\to \tau^+ \tau^-}}  \times  B_r\left(b\bar{b}\tau^+\tau^-\right)~.
\end{align}
We add the superscript ``LO'' to denote the leading-order expansion.
The NLO cross section can be expressed as a sum of two contributions, i.e., the NLO corrections to the production process combined with LO decays and the NLO corrections to the decay process interfaced with the LO production,
\begin{align}
   \int d \sigma_{\rm pro+dec}^{\rm \delta NLO} =&      \int d \sigma_{\rm pro+dec}^{ \rm \delta NLO^{pro} }   +  \int d \sigma_{\rm pro+dec}^{\rm  \delta NLO^{dec} }~,
\end{align}   
where
\begin{align}
    \int d \sigma_{\rm pro+dec}^{\rm \delta NLO^{pro} } =  \int d \sigma_{\rm pro}^{\rm  \delta NLO} 
    \times\frac{\int d\Gamma_{H_1\to b\bar{b}}^{\rm LO}}{\Gamma_{H\to b\bar{b}}^{\rm LO}} \frac{\int d\Gamma_{H_2\to \tau^+ \tau^-}}{\Gamma_{H\to \tau^+ \tau^-}}  \times  B_r\left(b\bar{b}\tau^+\tau^-\right)~,
\label{eq:nlopro}
\end{align}
and
\begin{align}
    \int d \sigma_{\rm pro+dec}^{\rm  \delta NLO^{dec}} & = \int d \sigma_{\rm pro}^{\rm  LO } \times  \frac{\int d\Gamma_{H_2\to \tau^+ \tau^-}}{\Gamma_{H\to \tau^+ \tau^-}}
 \frac{\int d\Gamma_{H_1\to b\bar{b}}^{\rm LO}}{\Gamma_{H\to b\bar{b}}^{\rm LO}}  \left( \frac{\int d\Gamma_{H_1\to b\bar{b}}^{\rm \delta NLO}}{\int d\Gamma_{H_1\to b\bar{b}}^{\rm LO}} -\frac{\Gamma_{H\to b\bar{b}}^{\rm \delta NLO}}{\Gamma_{H\to b\bar{b}}^{\rm LO}}  \right)  \times  B_r\left(b\bar{b}\tau^+\tau^-\right)~. 
     \label{eq:nlodec} 
\end{align}
The superscript ``$\delta {\rm NLO}$'' indicates the expansion at $\mathcal{O}(\alpha_s)$.
Note that we do not need to expand the decay width of $H\to \tau^+ \tau^-$ because it is not subject to NLO QCD corrections. 
Eq. (\ref{eq:nlopro}) represents the NLO QCD correction to the production process, 
while Eq. (\ref{eq:nlodec}) incorporates the NLO QCD correction to the decay process.
The second term in the bracket of Eq. (\ref{eq:nlodec}) comes from the QCD correction to the partial decay width $\Gamma_{H\to b\bar{b}}$ in the denominator of Eq. (\ref{eq:crxs}).
It will cancel with the first term if the latter is integrated over the full phase space.
In practice, multiple kinematic cuts are applied, and thus the decay process exhibits non-vanishing QCD corrections.
This is the reason why we keep the integration symbols $\int$ explicitly in the above equations.

Since ggF Higgs boson pair production is a loop-induced process in the SM, the virtual corrections involve two-loop calculations with multiple scales. 
In our calculation, we employ the numerical grid in {\tt POWHEG BOX}~\cite{Alioli:2010xd,Heinrich:2020ckp}.
The real corrections for $HH$ production come from one-loop diagrams such as $gg\to HHg$, $qg\to HHq$, $q\bar{q}\to HHg$. These contributions are computed using the package {\tt OpenLoops}~\cite{Buccioni:2019sur,Buccioni:2017yxi}, interfaced with {\tt Collier/OneLOop} to calculate the scalar integrals \cite{Denner:2016kdg,vanHameren:2010cp}.

All the amplitudes required for NLO QCD corrections to  $H\to b\bar{b}$  are computed using {\tt FeynArts}~\cite{KUBLBECK1990165,Hahn:2000kx} and {\tt FeynCalc}~\cite{MERTIG1991345,Shtabovenko:2016sxi,Shtabovenko:2020gxv} packages.
The analytic results can be found in our previous paper~\cite{Li:2024ujf}.  
In the calculation, we neglect the bottom-quark mass in the final state, which causes a correction only below $1\%$.
However, we use finite bottom-quark Yukawa coupling defined in the $\overline{\rm MS}$ scheme, which, in contrast to the on-shell mass, is insensitive to the non-perturbative effect.

Both virtual and real corrections contain infrared divergences, which are subtracted by using the integrated and differential dipole terms constructed following the method proposed in \cite{Catani:1996vz,Gleisberg:2007md}.
The implementation of the subtraction terms is checked by reproducing the known $gg\to HH$ production rate \cite{Borowka:2016ypz} and $H\to b\bar{b}$ decay width \cite{DelDuca:2015zqa} at NLO in QCD.

\section{Numerical results}\label{sec:num}

\begin{table*}[t]
\setstretch{1.2}
	\centering
	\begin{tabular}{|c|c|c|c|c|c|}
		\hline
		&without decays & \multicolumn{2}{c|}{with decays but no cuts}& \multicolumn{2}{c|}{with decays and cuts}\\ \cline{1-6}
		fb &  & ${\rm LO}^{\rm dec}$ & $\delta {\rm NLO}^{\rm dec}$ & ${\rm LO}^{\rm dec}$ & $\delta {\rm NLO}^{\rm dec} $  \\ 
  \cline{1-6}
		${\rm LO}_\infty^{\rm pro}$ &$15.72_{-22\%}^{+31\%}$ &$1.148_{-22\%}^{+31\%}$ &$0$&$0.6876^{+31\%}_{-22\%}$ &$-0.0924^{+42\%}_{-28\%}$ \\ \cline{1-1}
		${\rm LO}_{m_t}^{\rm pro}$ &$18.57^{+28\%}_{-20\%}$& $1.357^{+28\%}_{-21\%}$&$0$&$0.7765^{+27\%}_{-20\%}$&$-0.1361^{+40\%}_{-27\%}$ \\ \cline{1-1}
		$\delta {\rm NLO}_{\infty}^{\rm pro}$ & $13.68_{-7\%}^{+6\%}$&$0.9997^{+6\%}_{-7\%}$ & $-$&$0.5869^{+6\%}_{-7\%}$&$-$\\ \cline{1-1}
		$\delta {\rm NLO}_{m_t}^{\rm pro}$ & $12.27^{+4\%}_{-8\%}$ &$0.8964^{+4\%}_{-8\%}$ &$-$ &$0.5057^{+4\%}_{-8\%}$&$-$\\ \cline{1-6} 
            \multicolumn{6}{|c|}{Full NLO result} \\  \cline{1-6}
            ${\rm NLO}_{\infty}$ & $29.40^{+18\%}_{-15\%}$ & \multicolumn{2}{c|}{$2.148^{+18\%}_{-15\%}$}& \multicolumn{2}{c|}{$1.182^{+15\%}_{-14\%}$}\\ \cline{1-1}
            ${\rm NLO}_{m_t}$ & $30.84^{+14\%}_{-13\%}$ & \multicolumn{2}{c|}{$2.253^{+14\%}_{-13\%}$}& \multicolumn{2}{c|}{$1.146^{+10\%}_{-11\%}$}\\ \cline{1-6} 
	\end{tabular}
 \caption{Inclusive and fiducial cross sections for Higgs boson pair production and decay at the 13.6 TeV LHC. The LO and NLO results are provided with values computed with and without decay effects. The subscript $\infty$ denotes the cross section calculated in the heavy top-quark limit, while the subscript $m_t$ corresponds to the full SM result with complete top-quark mass dependence. The scale uncertainties are shown in percentage. The symbol `--' represents the higher-order corrections that we neglected in this work.  }
  \label{tab:totalxs}
\end{table*}

In numerical calculations, we adopt the following input parameters:
\begin{align}
& G_F = 1.1663787\times10^{-5} ~{\rm GeV}^{-2}~,
\quad m_W=80.399 ~{\rm GeV}~, \nn \\
& m_H = 125 ~{\rm GeV}~,  
~m_t = 173 ~{\rm GeV}~, 
~m_b(m_b) =4.18 ~{\rm GeV}~,\nn\\
&m_{\tau} = 1.777~{\rm GeV}~.
\end{align}
The Higgs boson decay branching ratios are taken to be $R(H\rightarrow b\bar{b})=0.5824$ and $R(H\rightarrow \tau^+\tau^-)=6.272\times10^{-2}$\cite{LHCHiggsCrossSectionWorkingGroup:2016ypw,Djouadi:2018xqq,Djouadi:1997yw}. We have used the parton distribution function set ${\rm PDF4LHC15\_nlo\_100\_pdfas}$ \cite{Buckley:2014ana} together with the associated strong coupling constant,  $\alpha_s$.  The renormalization and factorization scales are set by default to $\mu_r=\mu_f=m_{HH}/2$. 
The scale uncertainties are estimated by  varying $\mu_r$ and $\mu_f$ independently by a factor of two, excluding the cases of $\mu_r/\mu_f=4$ and $1/4$. 
The different choices of the top-quark mass renormalization scheme introduce another kind of uncertainty arising in the production process.
We find that this uncertainty almost remains the same after including the decay and kinematic cuts at LO.
The interested readers are referred to  \cite{Baglio:2018lrj,Baglio:2020ini,Bagnaschi:2023rbx} for the discussion on the scheme dependence at NLO for stable Higgs boson pair production.
To stabilize the numerical integration, we impose a technical cut on the Higgs boson pair transverse momentum, $p_T^{\rm min}=0.1~{\rm GeV}$, in the real corrections to the production, while a technical cut of  $s_{ij}^{\rm min} \sim 10^{-5} ~{\rm GeV^2}$ is applied to the real emissions in the decay.

In this work, we do not consider the decay of  $\tau$ leptons, since they can be reconstructed and identified using dedicated algorithms by the ATLAS~\cite{ATLAS:2022xzm,ATLAS:2014rzk} and CMS~\cite{CMS:2022hgz,CMS:2018jrd} collaborations. 
The $b$-jets are constructed using the anti-$k_t$ algorithm with a cone size $R=0.4$ as implemented in the {\tt FastJet} package~\cite{Cacciari:2011ma}.  
To investigate QCD corrections to the cross sections with kinematic cuts, we select events that satisfy the following criteria:
\begin{align}
   & p_T^j \ge 20 ~{\rm GeV}~, \quad
    p_T^{\tau} \ge 20 ~{\rm GeV}~, \quad
    |\eta^j| \le 2.4~, \quad
   \nn \\
&   |\eta^{\tau}| \le 2.3~, \quad 90~ {\rm GeV} \le m_{jj} \le 190 ~{\rm GeV}~, \quad \nn 
   \nn \\
&   R_{ll}>0.3~,\quad
    R_{jl}>0.5~,\quad
    R_{jj}>0.4~,\quad
\label{eq:cuts}
\end{align}
where $R_{ll}$ represents the distance between the two  $\tau$ leptons, $R_{lj}$ denotes the separation between a $b$-jet and a $\tau$ lepton, and $R_{jj}$ is the distance between two $b$-jets.

Table~\ref{tab:totalxs} shows the inclusive and fiducial cross sections for Higgs boson pair production and decay at the 13.6 TeV LHC.
We have performed the calculation in both the full SM and the heavy top-quark limit.
The results in the SM are higher than those in the heavy top-quark limit by 13\% and 18\% at LO with and without cuts on the decay products, respectively.
However,
the NLO QCD corrections in the production increase the cross section by 66\% and 87\% in the SM and large $m_t$ limit, respectively.
And the NLO QCD corrections in the decay decrease the cross section and are more pronounced in the SM, reaching $-18\%$.
Consequently, the NLO results after cuts in the SM are smaller than those in the heavy top-quark limit, although the difference is small, only 3\%.
The scale uncertainties are reduced by a factor of two in both the SM and large $m_t$ limit.

From the table, we observe that the cross sections with decays but without cuts are just the production rates multiplied by the branching ratio,
which serves as a check of our numerical program.
Note that the cuts in Eq.~(\ref{eq:cuts}) look rather loose.
However, they cut down over $40\%$ of the events at both LO and NLO.
One of the reasons is that the subleading $b$-jet and $\tau$ lepton do not have large transverse momentum in most of the events.

\begin{figure}[h!]
    \centering
    \includegraphics[width=0.32\linewidth]{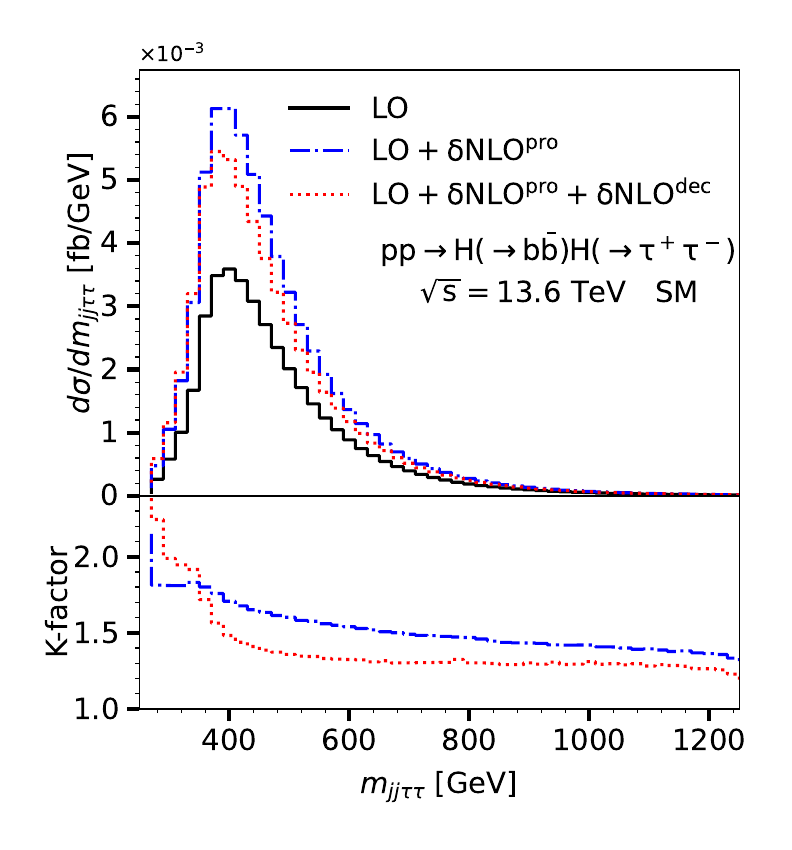}
    \includegraphics[width=0.32\linewidth]{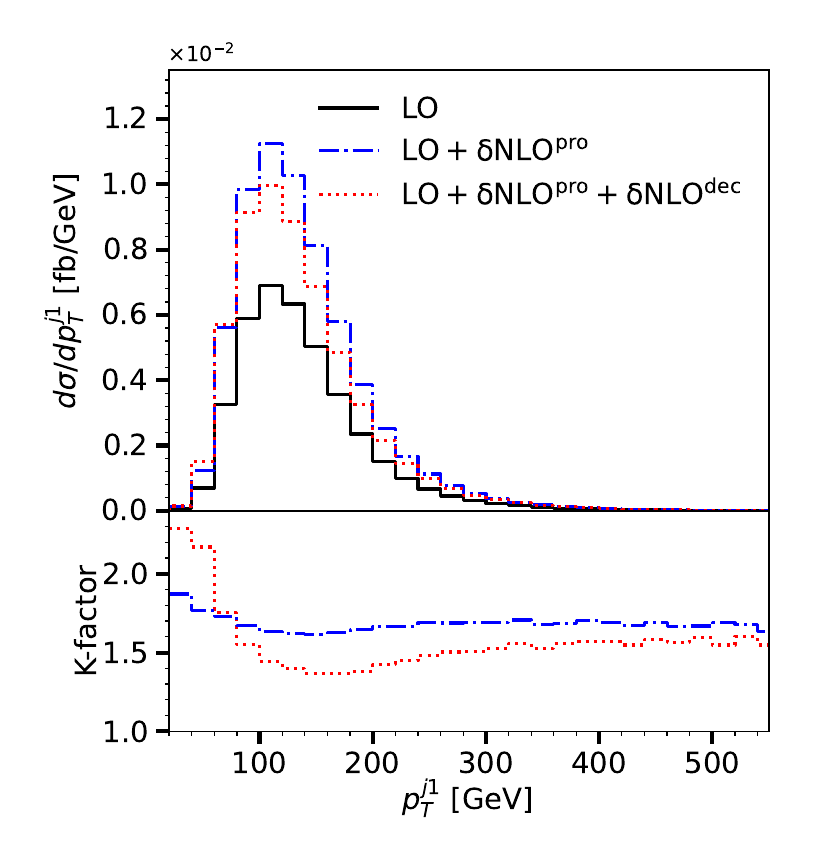}
    \includegraphics[width=0.32\linewidth]{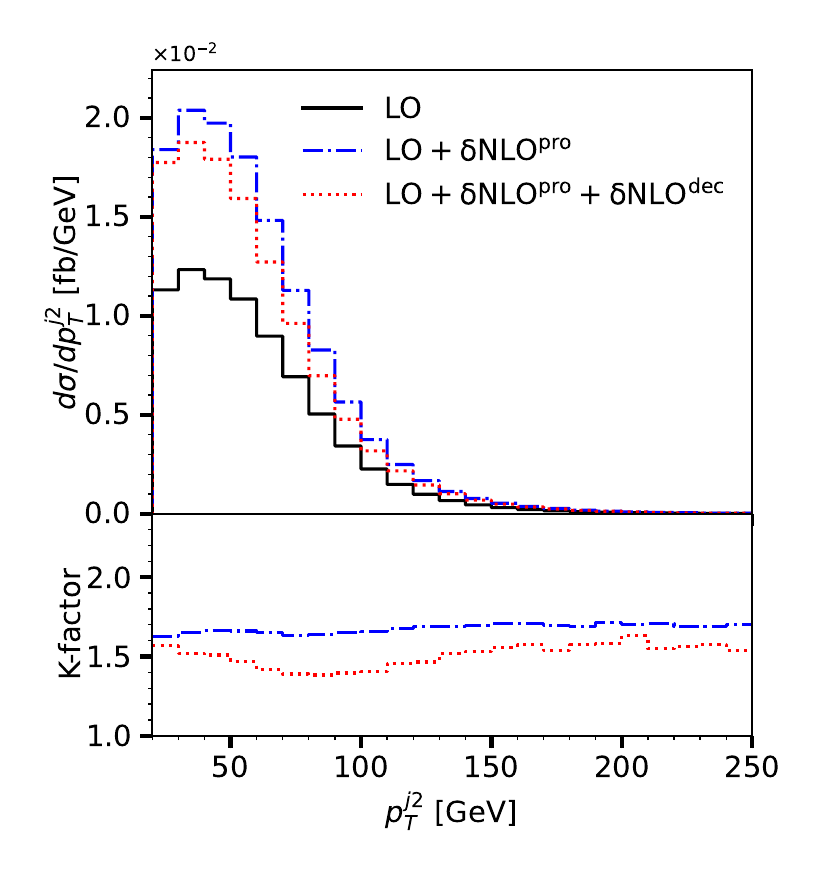}
    \includegraphics[width=0.32\linewidth]{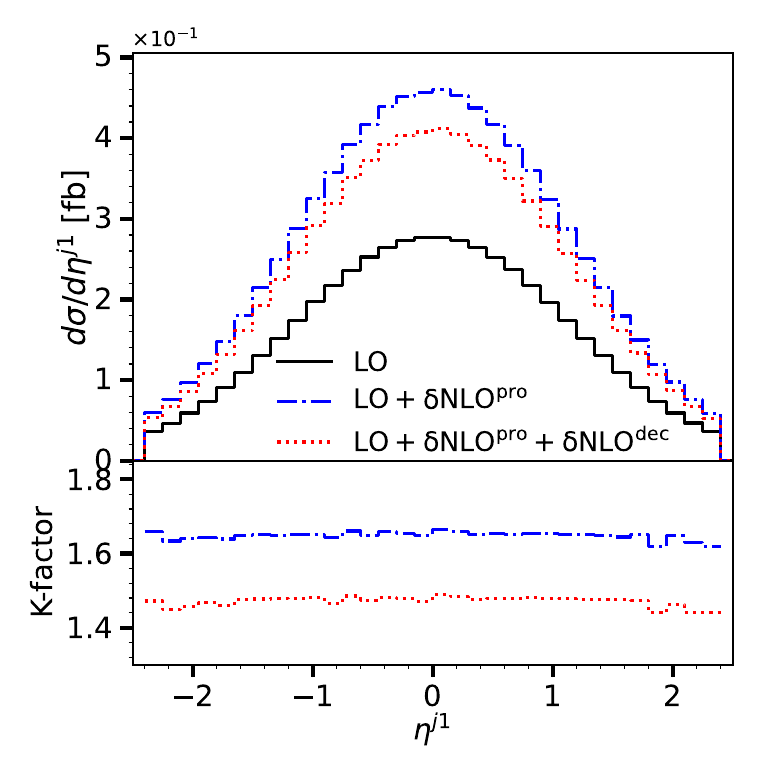}
    \includegraphics[width=0.32\linewidth]{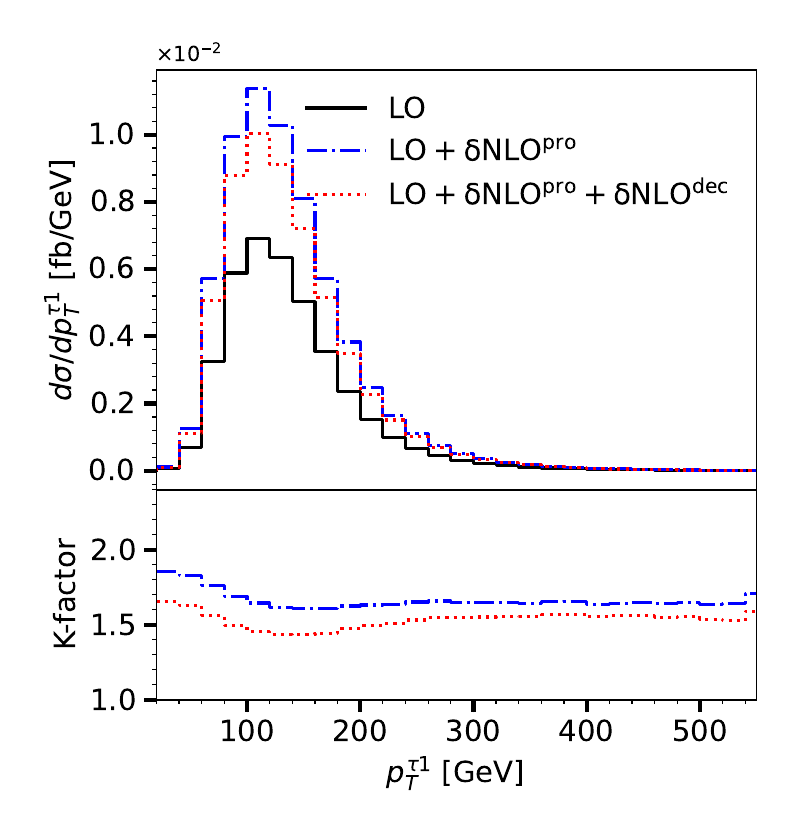}
    \includegraphics[width=0.32\linewidth]{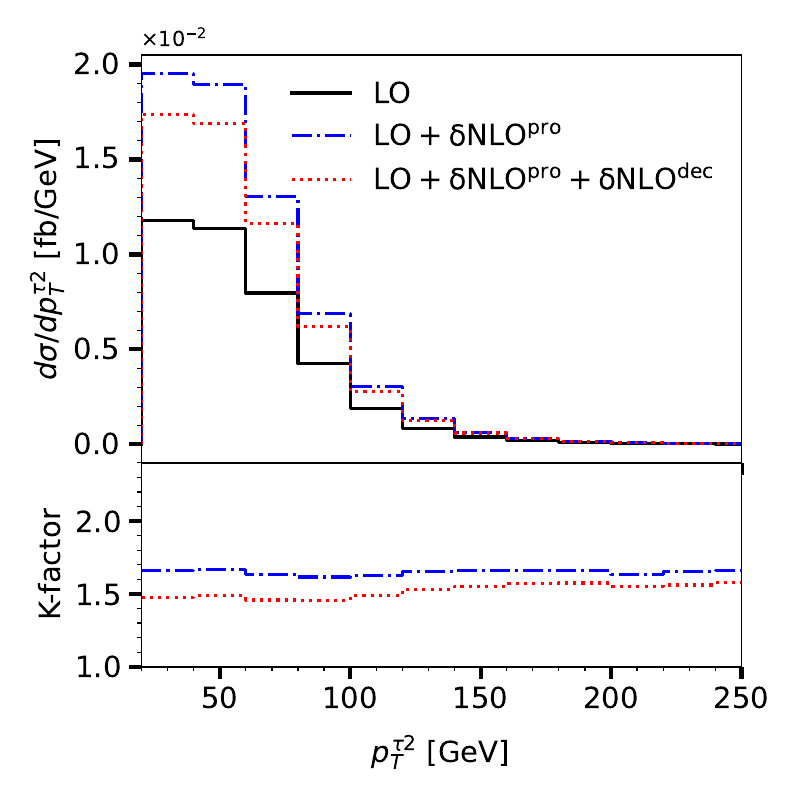}
    \vspace{0cm}
   \caption{
    Differential cross sections for Higgs boson pair production and decay with QCD corrections. The black curves represent the LO predictions, while the blue dashed curves show the results with NLO corrections to production only. The red dotted curves display the outcome when full QCD corrections (production and decay) are included. The K-factors are defined as the ratio of NLO results over LO ones.
   }
   \label{fig:kinematic_NLO_effect}
\end{figure}

In Fig.~\ref{fig:kinematic_NLO_effect}, we display the kinematic distributions of the final states with NLO QCD corrections after cuts.
In the  Higgs boson pair invariant mass $m_{jj\tau\tau}$ distribution, which is reconstructed from the final-state $b$-jets and $\tau$ leptons, 
the QCD corrections in production are more significant in the smaller $m_{jj\tau\tau}$ region, reaching 80\% at most.
The QCD corrections in decay increase and decrease the distribution below and above $m_{jj\tau\tau}=340$  GeV, respectively,
and therefore shift the peak towards a smaller value of $m_{jj\tau\tau}$.
In addition, they reduce the peak height by 11\%.

The $b$-jets are ordered according to their transverse momenta.
The shape of the leading $b$-jet transverse momentum distribution, denoted by $p_T^{j1}$, is notably changed after including the NLO QCD corrections in decay, since both positive and negative corrections exist in different regions of $p_T^{j1}$.
The subleading $b$-jet transverse momentum distribution is reduced by the QCD corrections in decay.
The effect is most obvious around $p_T^{j2}=80$ GeV.  The QCD corrections to the rapidity distribution of the $b$-jets and $\tau$ leptons are almost flat over the whole region. We show the case of the leading $b$-jet in Fig.~\ref{fig:kinematic_NLO_effect}.

The transverse momentum distributions of the $\tau$ leptons, shown in Fig.~\ref{fig:kinematic_NLO_effect}, also suffer from the suppression of QCD corrections in decay, especially in the peak regions.

\begin{figure}[h!]
    \centering    
    \includegraphics[width=0.32\linewidth]{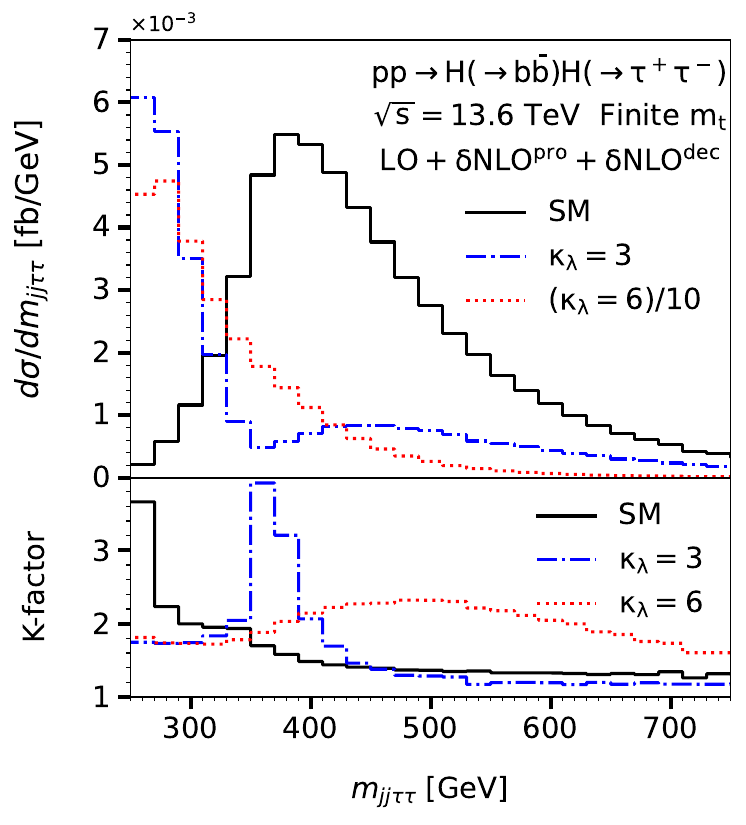}
    \includegraphics[width=0.32\linewidth]{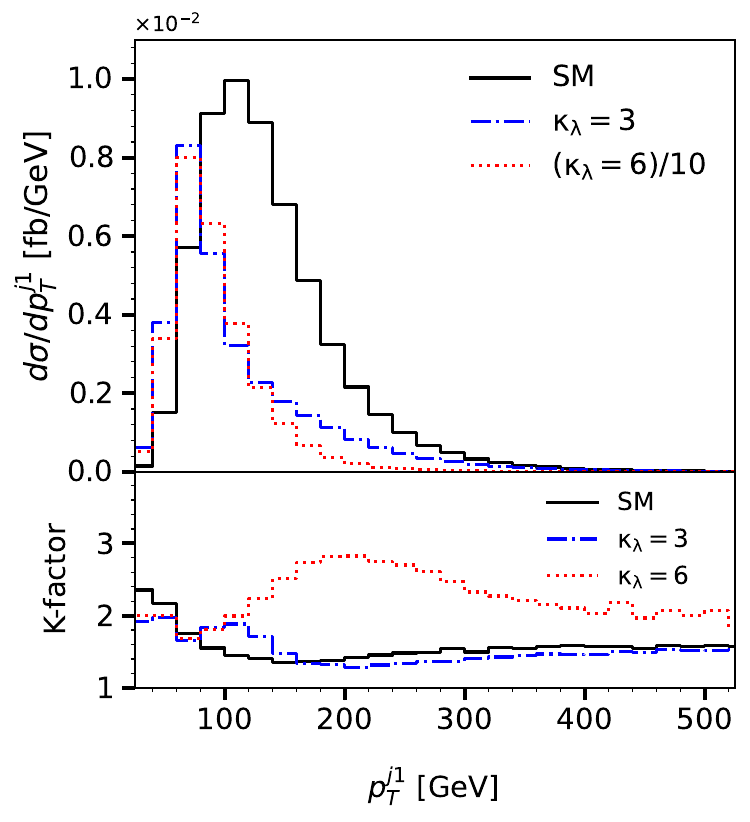}
    \includegraphics[width=0.32\linewidth]{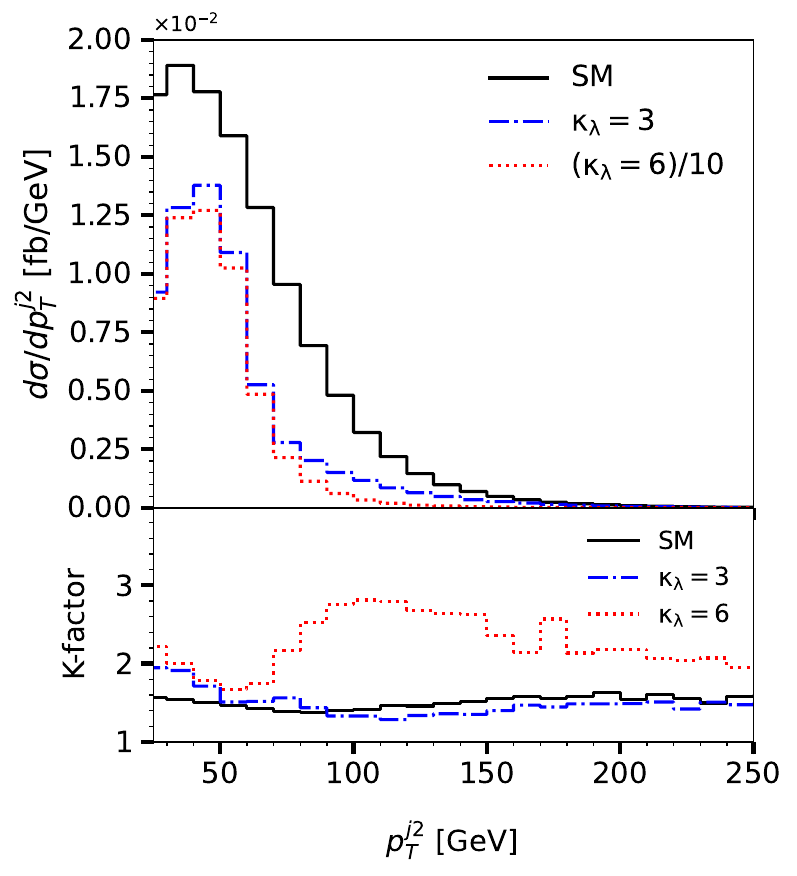}
    \includegraphics[width=0.32\linewidth]{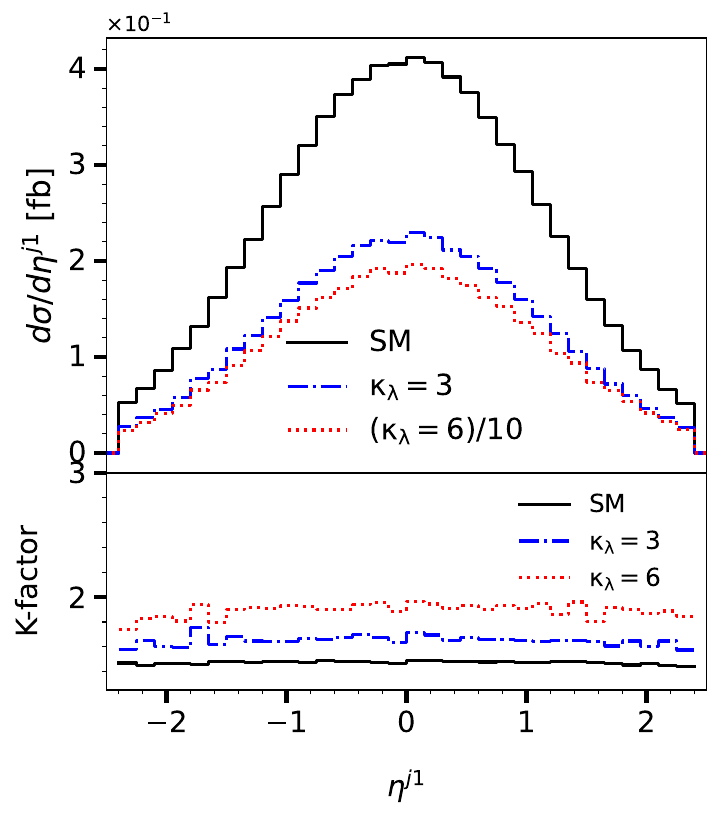}
    \includegraphics[width=0.32\linewidth]{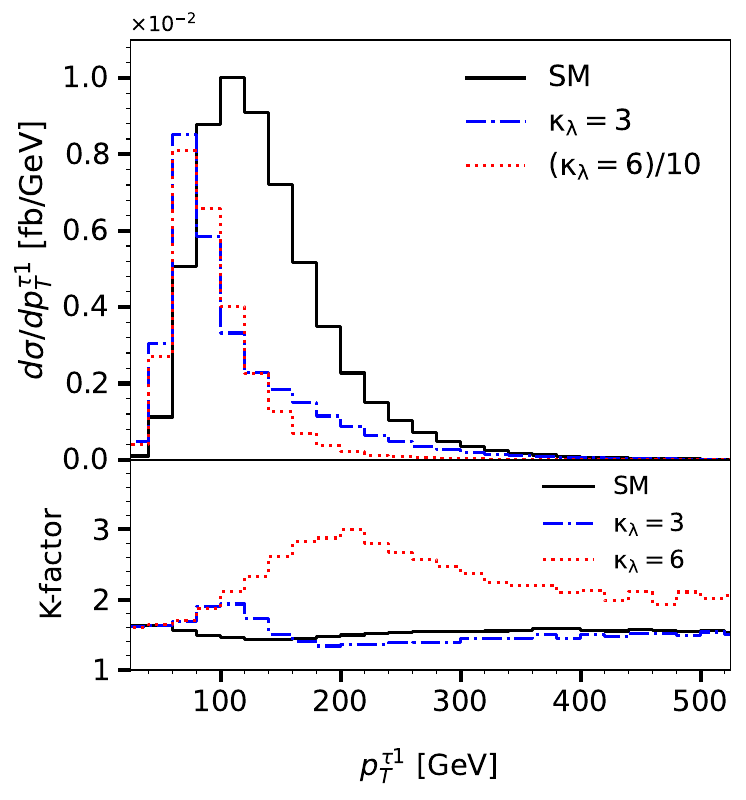}
    \includegraphics[width=0.32\linewidth]{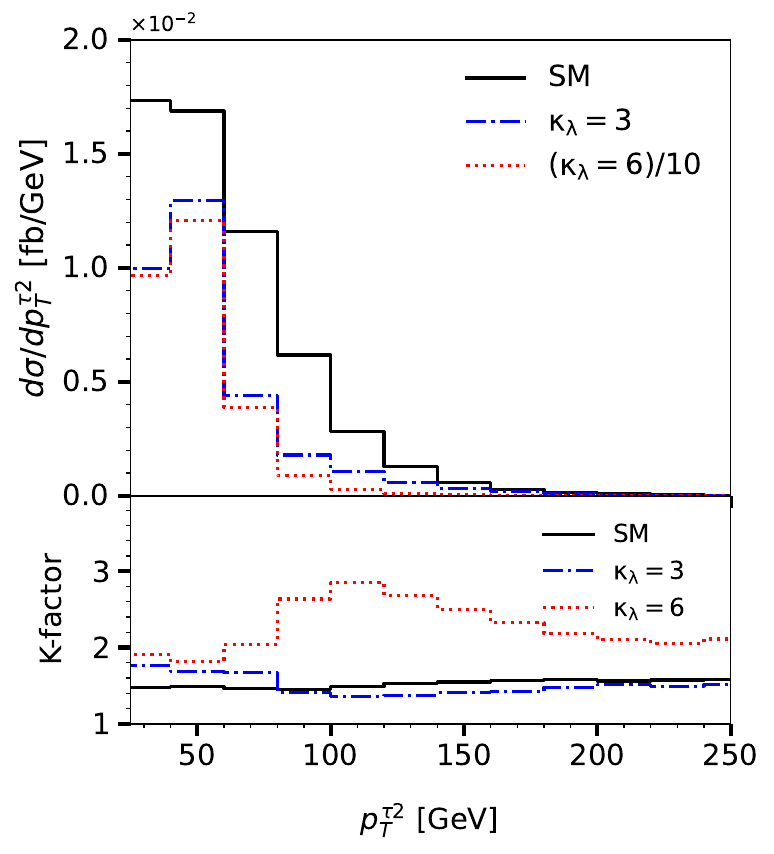}
   \vspace{0cm}
   \caption{Differential cross sections for Higgs boson pair production and decay at different values of $\kappa_\lambda$.
    The results of $\kappa_\lambda=6$ have been rescaled by a factor of $1/10$.
   The lower panels show the K-factors defined as NLO QCD results over the LO ones.}
   \label{fig:compared_kappa}
\end{figure}

The kinematics of the signal events at different $\kappa_{\lambda}$ plays a crucial role in determining  the experimental constraints on the Higgs self-coupling,
since the signal acceptance, defined as the cross section passing the cuts over the {\it total} cross section, is generally not a constant for different $\kappa_{\lambda}$ \cite{ATLAS:2019ocm,ATLAS:2022jtk}.
We show in Fig. \ref{fig:compared_kappa} the kinematics at different values of $\kappa_{\lambda}$. The lower panels in Fig. \ref{fig:compared_kappa}  show that the QCD corrections are important to provide precise theoretical predictions for various distributions and do not appear as overall enhancement factors. 
It can be seen that the invariant mass $m_{jj\tau\tau}$ distribution varies dramatically as $\kappa_{\lambda}$ changes from 1 to 6.
Near the threshold of Higgs boson pair production (around $250$ GeV),
the SM cross section is almost vanishing due to the cancellation between the amplitudes with and without the Higgs self-coupling dependence.
Choosing larger values of $\kappa_{\lambda}$ breaks the cancellation and the cross section is remarkably enhanced in this region.
Then the cancellation would happen in other regions of $m_{jj\tau\tau}$.
This explains the dip around $m_{jj\tau\tau}=360$ GeV in the curve for $\kappa_{\lambda}=3$.
Whenever the cancellation happens,
QCD corrections increase the cross section tremendously,
which can be observed by the large K-factor (larger than 3.6) in the lower panel.

The transverse momentum distributions of the $b$-jets and $\tau$ leptons at different values of $\kappa_{\lambda}$ have very different shapes.
The leading $b$-jets and leptons tend to be softer as the increasing of $\kappa_{\lambda}$,
while the peak positions of the subleading $b$-jets and leptons move towards harder regions. 
In all these distributions, prominent QCD corrections can be seen, in particular, for large $\kappa_{\lambda}$.

Since the (differential) cross section is a quadratic function of $\kappa_{\lambda}$, the distributions for other values of $\kappa_{\lambda}$ can be derived from the three curves in the plots, which correspond to $\kappa_{\lambda}=1,3,$ and 6, respectively.

\begin{figure}[!h]
    \centering    
    \includegraphics[width=0.5\linewidth]{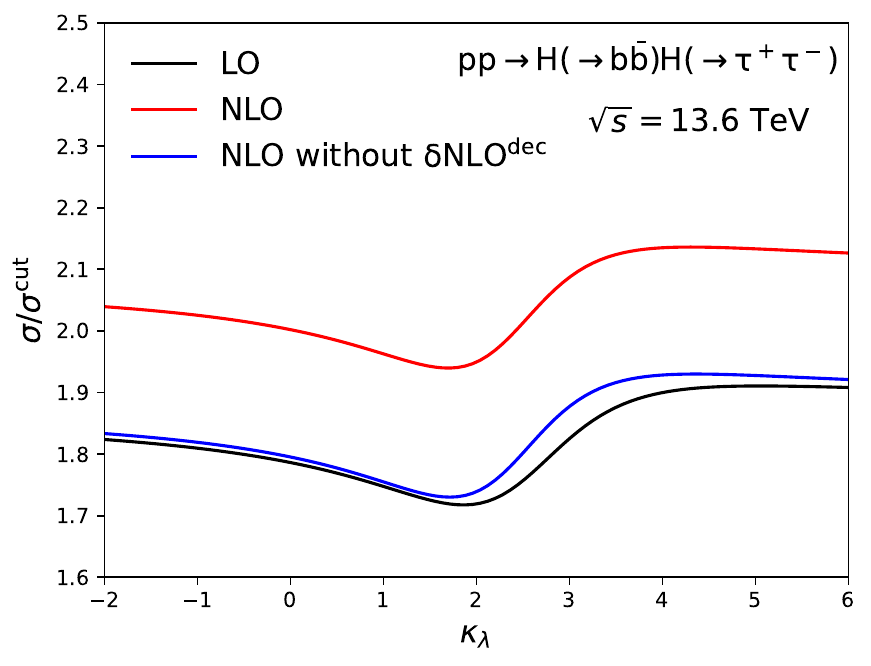}
   \vspace{0cm}
   \caption{The inverse of signal acceptance as a function of $\kappa_\lambda$.}
   \label{fig:kappa_ratio}
\end{figure}

Based on the above calculation,
 we obtain
the cross sections  after the cuts in Eq. (\ref{eq:cuts}) as functions of $\kappa_{\lambda}$,
\begin{align}
    \sigma^{\rm cut}_{\rm LO} &= 0.2028\kappa_{\lambda}^2-1.010\kappa_\lambda+1.583 ~{\rm fb} \,,\nonumber \\
    \sigma^{\rm cut}_{\rm NLO} &= 0.3703\kappa_\lambda^2-1.740\kappa_\lambda+2.516 ~{\rm fb}\,, \nonumber \\
    \sigma^{\rm cut}_{\rm NLO / \delta NLO^{dec}} &= 0.4107\kappa_\lambda^2-1.934\kappa_\lambda+2.806 ~{\rm fb}\,,
\end{align}
where the third equation represents the result without QCD corrections in decay.
Comparing them to the total cross sections without cuts
\begin{align}
    \sigma^{}_{\rm LO} &= 0.3797\kappa_{\lambda}^2-1.852\kappa_\lambda+2.829~{\rm fb} \,, \nonumber \\
    \sigma_{\rm NLO} &= 0.7728\kappa_\lambda^2-3.560\kappa_\lambda+5.037 ~{\rm fb}\,,
\end{align}
we see that a large amount of the events cannot pass the cuts for a general $\kappa_{\lambda}$.
The inverse of the signal acceptance, i.e., $\sigma/\sigma^{\rm cut}$,
multiplied with the cross section measured experimentally, 
provides valuable information on the upper bound of the {\it total} cross section of $HH$ production.
We depict  in Fig.~\ref{fig:kappa_ratio}  the inverse of the signal acceptance under the cuts in Eq.~(\ref{eq:cuts}).
A dip occurs near $\kappa_{\lambda}=2$, 
which is a feature also in the plot on the expected exclusion limit in the ATLAS analysis \cite{ATLAS:2022jtk}.
From Fig.~\ref{fig:kappa_ratio}, the allowed region of $\kappa_{\lambda}$ would become wider after considering the QCD NLO corrections, especially those in the decay process.

\section{Conclusion}
\label{sec:concl}

Higgs boson pair production via gluon-gluon fusion provides a unique probe of the Higgs self-coupling, a critical parameter to test the SM. 
In this work, we have presented a precise theoretical prediction for this process and the subsequent decay to 
$b\bar{b}\tau^+\tau^-$, which possesses a large branching fraction and distinctive kinematic features.
Our study reveals that, once typical experimental cuts are applied, the NLO QCD corrections to the decay process cause a reduction of approximately 18\% in the fiducial cross section compared to the LO prediction. Moreover, the kinematic distributions, particularly those of the invariant mass $m_{jj\tau\tau}$ and the leading 
$b$-jet transverse momentum, are remarkably reshaped by these corrections. We have further demonstrated that variations in the Higgs self-coupling modifier $\kappa_\lambda$  lead to significant shifts in the kinematic distributions, underscoring the sensitivity of the  $b\bar{b}\tau^+\tau^-$ channel to new physics effects,
and that the QCD corrections are more pronounced at larger $\kappa_{\lambda}$.
These results demonstrate the necessity of including full QCD corrections in theoretical predictions to ensure precise extraction of the Higgs self-coupling from current and future experimental data.

\section*{Acknowledgements}
This work was supported by the National Natural Science Foundation of China under grant No. 12275156, No. 12321005, No. 12375076 and the Taishan Scholar Foundation of Shandong province (tsqn201909011). 

\bibliographystyle{JHEP} 
\bibliography{ref.bib}

\end{document}